\newtheorem{cor}{Corollary}
\newtheorem{prop}{Proposition}
\newtheorem{theo}{Theorem}

\documentstyle[tighten,aps,amstex,amscd]{revtex}

\newcommand{\Rn}{{\rm I\!R}} 
\newcommand{\Cn}{{\setbox0=\hbox{
$\displaystyle\rm C$}\hbox{\hbox
to0pt{\kern0.6\wd0\vrule height0.9\ht0\hss}\box0}}} 
\newcommand{\Zn}{{\hbox{$\sf\textstyle Z\kern-0.3em Z$}}} 

\newcommand{\cA}{{\cal A}}
\newcommand{\cB}{{\cal B}}

\newcommand{\cE}{{\cal E}}
\newcommand{\cF}{{\cal F}}
\newcommand{\cH}{{\cal H}}
\newcommand{\cI}{{\cal I}}

\newcommand{\cM}{{\cal M}}
\newcommand{\cN}{{\cal N}}

\newcommand{\cO}{{\cal O}}

\newcommand{\cS}{{\cal S}}

\newcommand{\cZ}{{\cal Z}}

\begin{document}
\twocolumn
\makeatletter
\title{On the measure of entanglement}

\author{Adam W. Majewski}

\address{Institute of Theoretical Physics and Astrophysics, University
of Gda\'nsk, Wita Stwosza 57, PL 80-952 Gda\'nsk, Poland.  E-mail: fizwam@univ.gda.pl}

\date{\today}
\maketitle

\begin{abstract}
In this paper we present the novel qualities of entanglement
of formation for general (so also infinite dimensional)
 quantum systems. A major benefit of our
presentation is a rigorous description of entanglement of formation. 
In particular, we indicate how this description may be used to examine 
optimal decompositions.
Illustrative examples showing the method
of estimation of entanglement of formation are given.
\end{abstract} 


\section{Introduction}
The problem of quantum entanglement of mixed states has attracted
much attention recently and it has been widely considered in different 
physical contexts (cf.  \cite{Hor1} and references therein, 
see also \cite{KM}). Due to recent
works by Peres \cite{per} and Horodeccy \cite{Hor1}, 
\cite{hor} there exists a 
simple criterion allowing one to judge whether a given density matrix 
$\varrho$ representing a $2 \times 2$ or $2 \times 3$
composite system, is separable. On the other hand, the definition
of a measure of entanglement for general quantum systems 
as well as the problem 
of finding operational, sufficient and necessary
condition for separability in higher dimensions remain still open
(cf. \cite{kus},\cite{lew}, and \cite{Hor1} and references therein).

In this paper we are concerned with the 
entanglement of formation, EoF, introduced in \cite{Ben}.
Let us stress that the principal motivation for our generalization 
of the definition of EoF 
follows from
foundations of Quantum Mechanics; a quantum system is described
by infinite dimensional Hilbert space.
Further, to indicate that this concept stems from 
mathematical structure of tensor product we develop the theory
of entanglement of formation in general terms of composite systems.
Moreover, we look more closely at the original definition of EoF. Namely,
there is a difficulty in implementing the definition given by 
Bennett {\it et al}
in the sense that it is not clear why
the operation of taking $min$ over the set of
all decomposition of the given state into 
finite convex combination of pure states
is well defined (for details see subsection
{\it Optimal decompositions} is Section V). To overcome this problem
and to have a measure with nice topological properties  we shall
use the theory of decomposition which is based on the theory
of compact convex sets and boundary integrals.
The paper is organized as follows. In Section II we set up notation
and terminology, and we review some of the standard facts on the theory
of decomposition. Section III contains the definition of entanglement of 
formation, EoF, with the proof that {\it EoF is equal to zero if and
only if a state is a separable one}. In other words, EoF, leads to well 
established criterion of separability.
In section IV we review properties of EoF.
Namely, we indicate how techniques based on decomposition theory 
can be used to study EoF. Moreover, some simple examples showing the method
of estimation of EoF are given.
Furthermore, the proof of convexity and detailed study of 
topological properties 
of EoF are obtained. In particular, it is shown that {\it the family
of maximally entangled states is a subset of pure states}.
In the final section V, we present some
other examples of explicitly calculated EoF and we clarify the relation
between our and the Bennett's {\it et al} definition.
Furthermore, we provide a detailed exposition of the concept of optimal
decompositions. Also, some 
remarks concerning uniqueness of measure of entanglement are given.

\section{Preliminaries}

Let us consider a composite system $"1+2"$ and its Hilbert space
of the pure states ${\cal H}_1 \otimes {\cal H}_2$ 
where ${\cal H}_i$ is the Hilbert space associated to subsystem
$i$ ($i = 1,2$). Let ${\cal B}({\cal H}_1) $,
denote the set of all bounded
linear operators on ${\cal H}_1$ while $\cM$ stands for a
(unital) $C^*$-subalgebra of 
$\cB(\cH_2)$.
We will assume that $\cH_1$ is finite dimensional
space (in concluding remarks, Section V, we will indicate how to dispense
with that assumption).
$\cH_2$ will be an arbitrary (infinite dimensional,
separable) Hilbert space.
In other words, the composite system consists of small subsystem and 
a big heat-bath, rather typical situation for concrete physical
problems. 
Any density matrix (normal state) on $\cal H$ determines uniquely a linear
positive, normalized, functional $\omega_{\varrho}(\cdot) \equiv \omega (\cdot)
\equiv Tr\{ \varrho \cdot \}$ on ${\cal B}({\cal H})$ which is also called 
a state.

We will assume the Ruelle's separability condition for $\cM$ (cf. \cite{Ru1},
\cite{Ru2}, \cite{BR}): a subset $\cF$ of the set of all states $\cS$ 
of $\cM$ satisfies separability condition if there exists a sequence
$\{ \cM_n \}$ of sub-$C^*$-algebras of $\cM$ such that 
$\cup_{n \ge 1} \cM_n$ is dense in $\cM$, and each $\cM_n$ contains
a closed, two-sided, separable ideal $\cI_n$ such that
\begin{equation}
\cF = \{ \omega; \omega \in \cS, ||\omega|_{\cI_n} || = 1, n\ge 1 \}
\end{equation}

This condition leads to a situation in which the subsets of states have good measurability properties. Furthermore, one can verify that this separability
condition is satisfied for two important cases:
\begin{itemize}
\item{}$\cM$ is a separable $C^*$-algebra. Then $\cS$ is metrizable
and Borel and the Baire structures on $\cS$ coincide. 
We put $\cF = \cS$ in that case.
\item{}$\cM = \cB(\cH)$ for some Hilbert space $\cH$ and $\cF$
is the set of all density matrices (normal states).
\end{itemize}
Thus, the separability condition covers the basic models of quantum mechanics
and we will restrict our attention to models satisfying this condition.
However, generalizations of our approach are possible.

We recall that the density matrix $\varrho$ (state) on the Hilbert space 
${\cal H}_1 \otimes {\cal H}_2$ is called separable if it can be written or 
approximated (in the norm) by the density matrices (states)
of the form: 
$$\varrho = \sum p_i \varrho_i^1 \otimes \varrho_i^2 \qquad 
\Biggl( \omega(\cdot)
= \sum p_i (\omega^1_i \otimes \omega^2_i) ( \cdot) \Biggr)$$
where $p_i \ge 0$, $\sum_i p_i = 1$, $\varrho^{\alpha}_i$ are density
matrices on ${\cal H}_{\alpha}$, $\alpha = 1,2$, and
$(\omega_i^1 \otimes \omega_i^2)(A \otimes B)
\equiv \omega^1_i(A) \cdot \omega^2_i(B) \equiv (Tr \varrho_i^1 A) 
\cdot (Tr \varrho_i^2 B) \equiv Tr\{ \varrho_i^1 \otimes \varrho_i^2 \cdot
A \otimes B\}$.

\smallskip

Now, for the convenience of the reader, we introduce some terminology
and give a short resum\'e of results from convexity and Choquet
theory that we shall need in the sequel
(for details see \cite{phel}, \cite{skau}, \cite{Mey}, and \cite{BR}). Let 
$\cA$ stand for a $C^*$-algebra. From now on we make the same assumption
of separability for $\cA$ which was posed for $\cM$. In next sections, by a 
slight abuse of notation we will write $\cA$ for $\cB(\cH_1) \otimes \cM$.
By $\cS$ we will denote the state space of 
$\cA$, i.e. the set of linear, positive, normalized,
 linear functionals on $\cA$.
We recall that $\cS$ is a compact convex set in the $^*$-weak topology.
Further, we denote by $M_1(\cS)$ the set of all probability
Radon measures on $\cS$. It is well known that
$M_1(\cS)$ is a compact subset of the vector space of real, regular
Borel measures on $\cS$. After these preliminaries let us recall
the concept of barycenter $b(\mu)$ of a measure $\mu \in M_1(\cS)$:
\begin{equation}
b(\mu) = \int d\mu (\varphi) \varphi
\end{equation}
where the integral is understood in the weak sense. The set
$M_{\omega}(\cS)$ is defined as a subset of $M_1(\cS)$
with barycenter $\omega$, i.e.
\begin{equation}
M_{\omega}(\cS) = \{ \mu \in M_1(\cS), b(\mu) = \omega \}
\end{equation}
$M_{\omega}(\cS)$ is a convex closed subset of $M_1(\cS)$, hence
compact in the weak $^*$-topology.
Hence, it follows by the Krein-Milman theorem that there are "many"
 extreme points in $M_{\omega}(\cS)$. We say 
the measure $\mu$ is simplicial
if $\mu$ is an extreme point in $M_{\omega}(\cS)$. We denote by 
$\cE_{\omega}(\cS)$ the set of all simplicial measures in $M_{\omega}(\cS)$.
Finally, we will need the concept of orthogonal measures.
To define that concept one introduces firstly the notion of orthogonality
of positive linear functionals on $\cA$: given
positive functionals $\phi, \psi$ on $\cA$
we say that $\phi$ and $\psi$ are othogonal, in symbols, $\phi \bot \psi$,
if for all positive linear functionals $\gamma$ on $\cA$, $\gamma \le \phi$
and $\gamma \le \psi$ imply that $\gamma = 0$.

Turning to measures, let $\mu$ be a regular non-negative
Borel measure on $\cS$ and let $\mu_V$ denote the restriction
of $\mu$ to $V$ for a measurable set $V$ in $\cS$, i.e.
$\mu_V(T) = \mu(V \cap T)$ for $T$ measurable in $\cS$.
If for all Borel sets $V$ in $\cS$ we have
\begin{equation}
\int_{\cS} \varphi d\mu_V(\varphi) \quad \bot 
\quad \int_{\cS} \varphi d\mu_{\cS \setminus V}(\varphi)
\end{equation}
we say that $\mu$ is an orthogonal measure on $\cS$. We recall
that the set of all othogonal measures on $\cS$
with barycenter $\omega$, $O_{\omega}(\cS)$, forms a subset
(in general proper) of $\cE_{\omega}(\cS)$, i.e.
$O_{\omega}(\cS) \subset \cE_{\omega}(\cS)$.

\section{Entanglement of Formation}

Let us define, for a state $\omega$ on ${\cal B}({\cal H}_1) \otimes
\cM$ the following map:
\begin{equation}
\label{r}
(r \omega)(A) \equiv \omega(A \otimes {\bf 1})
\end{equation}
where $A \in {\cal B}({\cal H}_1)$.

Clearly, $r \omega$ is a state on ${\cal B}({\cal H}_1)$. One has

\smallskip

{\it Let $(r \omega)$ be a pure state on ${\cal B}({\cal H}_1)$
(so a state determined by a vector
from ${\cal H}_1$). Then $\omega$ can be written as a product state on
${\cal B}({\cal H}_1) \otimes \cM$.}

The proof of that statement can be extracted from \cite{tak}. However, for the 
convenience of the reader we provide the basic idea of the proof.
It is enough to consider the case with an arbitrary but fixed
positive $B$ in unit ball of $\cM$ such that
$0< \omega ({\bf 1} \otimes B) < 1$. Then $(r \omega)(A)$ can be written
as
\begin{equation}
(r \omega)(A) = \omega({\bf 1} \otimes B) \omega^I(A) + (1 - 
\omega({\bf 1} \otimes B)) \omega^{II}(A)
\end{equation}
where $\omega^I(A) = {1 \over {\omega({\bf 1} \otimes B)} } 
\omega(A \otimes B)$ and $\omega^{II}(A) =
{1 \over {1 - \omega({\bf 1} \otimes B)}} \omega(A \otimes( {\bf 1}- B))$.
Clearly $\omega^I$ and $\omega^{II}$ are well defined linear,
positive functionals (states) on ${\cal B}({\cal H}_1)$. Hence, the purity
of $(r \omega)$ implies $\omega^I = \omega^{II}$. Consequently,
$\omega(A \otimes B) =
\omega(A \otimes {\bf 1}) \omega({\bf 1} \otimes B)$.
The rest is straighforward so the proof is completed.
(For more details we refer the reader to \cite{tak}, \cite{MM}).

Conversely, there is another result in operator 
algebras saying that if $\omega$ 
is a state on ${\cal B}({\cal H}_1)$ then there exists a state
$\omega^{\prime}$ over ${\cal B}({\cal H}_1) \otimes \cM$
which extends $\omega$. If $\omega$ is a pure state of
${\cal B}({\cal H}_1)$ then $\omega^{\prime}$ may be chosen
to be a pure state of ${\cal B}({\cal H}_1) \otimes \cM$
(cf. \cite{BR}).

\bigskip
Now we are in position to give a modification and discuss
the definition of entanglement of formation (cf. \cite{Ben}).
\bigskip

Let $\omega$ be a state on ${\cal B}({\cal H}_1) \otimes \cM$. 
{\it The entanglement of formation, EoF, is
 defined
as
\begin{equation}
{E}(\omega) = inf_{\mu \in M_{\omega}(\cS)}
\int_{\cS} d\mu(\varphi) S(r\varphi)
\end{equation}
where $S(\cdot)$ stands for the von Neumann entropy, i.e. $S(\varphi)
= - Tr \varrho_{\varphi} log \varrho_{\varphi}$
 where $\varrho_{\varphi}$
is the density matrix determining the state $\varphi$.}
\vskip 1cm

In order to comment the above definition we recall that the map $r$ 
and the function $S$ are ($^*$-weakly ) continuous.
At this point we want to strongly emphasize that we use 
the entropy function $S$ only to respect the tradition. Namely, to have
a well defined concept of EoF we need a concave 
non-negative continuous function 
which vanishes on pure states (and only on pure states).
In our case, with the first subsysten being finite, the von Neumann entropy
meets these conditions.
Thus, we define EoF as infimum of integrals evaluated on continuous
function and the infimum is taken over the compact set. Therefore,
the infimum is attainable, i.e. there exists a measure
$\mu_0 \in M_{\omega}(\cS)$ such that
\begin{equation}
E(\omega) = \int_{\cS} d\mu_0(\varphi)S(r \varphi)
\end{equation}
and
\begin{equation}
\omega = \int_{\cS} d\mu_0(\varphi)\varphi
\end{equation}

\vskip 1cm 

Now we want to show that $\cF \ni \omega \mapsto E(\omega) $ 
is equal to $0$ 
only for separable states (we recall that $\cF$ stands for the subset of states satisfying Ruelle's condition, cf. Section II). Assume $E(\omega) = 0$. Then
\begin{equation}
\int_{\cS} d\mu_0(\varphi) S(r\varphi) = 0, 
\end{equation}
for some probability measure $\mu_0$. As $S(r \varphi) \ge 0$
and it is the continuous function we infer that $S(r \varphi) = 0$
for each $\varphi$ in the support of $\mu_0$.
But, as the entropy is a concave function we have
\begin{equation}
S\circ r ( \varphi) \ge \int d\xi(\nu) S(r \nu)
\end{equation}
for any positive measure $d\xi$ on $\cS$ such that $\varphi = \int d\xi(\nu) \nu$. In particular, taking a measure supported on pure states (such decomposition always exists under the assumed
separability condition) we infer $S(r \nu) = 0$
so $r\nu$ is a pure state and consequently
$\nu$ is a product state. So $\varphi $ is a convex combination
of product states. Finally, as $\omega = \int_{\cS} d\mu_0 (\varphi) \varphi$
and $\mu_0$ can be well approximated by finite measures 
(see \cite{Bor})we infer that
$\omega$ can be approximated by convex combinations 
of product states, so $\omega$ is a separable state. 

Now, let us assume that $\omega$ is a separable state, i.e.
$\omega$ can be approximated by convex combinations of product states
$\omega_i^{(N)}$:
\begin{equation}
\omega = lim_N \sum_{i=1}^N \lambda_i^{(N)} \omega_i^{(N)}
\end{equation}
Define
\begin{equation}
\mu^N = \sum_{i = 1}^N \lambda_i^{(N)} \delta_{\omega_i^{(N)}}
\end{equation}
where $\delta_{\omega_I^{(N)}}$ are the Dirac measures of the point 
$\omega_i^{(N)}$. Considering the weak limit
of $\int d\mu^N(\varphi) \varphi$ we can infer that there is a measure
$\mu$ such that 
\begin{equation}
\int d\mu(\varphi) \varphi = \omega, \qquad \int d\mu(\varphi) 
S(r \varphi) = 0.
\end{equation}
 So we arrived to
\begin{theo}
A state $\omega \in \cF$ is separable if and only if EoF $E(\omega)$
is equal to 0.
\end{theo}

\section{Properties of EoF}

\subsection{Relations to other decompositions}

Let us discuss some relations between decompositions used
in our definition of EoF and other types of decompositions.
Assume that the state $\omega$ is separable, so
there is a measure $\mu_0 \in M_{\omega}(\cS)$
 such that $\int d\mu_0(\varphi)
S(r \varphi) =0$. But as we consider non-negative function, and
positive measures this implies that there is a simplicial measure
$\mu_0^s$
(in fact there can be many such measures) such that
$\int d\mu_0^s(\varphi) S(r \varphi) =0$.
In other words, the infimum is attainable on the set of simplicial measures
$\cE_{\omega}(\cS)$ (for more detailed discussion on the role of simplicial
measures in the description of EoF see subsection {\it Opitimal decompositions}
in Section V).

On the other hand, as $O_{\omega}(\cS) 
\subset \cE_{\omega}
(\cS)$ we have
\begin{equation}
\inf_{\mu \in \cE_{\omega}(\cS)}
\int d\mu(\varphi) S(r \varphi) \le \inf_{\mu \in O_{\omega}(\cS)}
\int d\mu(\varphi) S(r \varphi)
\end{equation}

In general we can not expect the equality in $(15)$.
 Namely, there are examples
of simplicial measures which are not orthogonal
(cf. \cite{Ef}). So finding an othogonal
 measure such that ``inf'' is attained we can 
infer that the state is separable but not conversely. 
To be more clear, let us recall some
algebraic aspects of decomposition theory (cf. \cite{BR}) which are 
related to orthogonal measures. A finite convex decomposition
of $\omega \in \cS$ corresponds to a finite decomposition
of identity ${\bf 1} = \sum_i T_i$, $T_i \ge 0$ within the commutant
$\pi_{\omega}(\cA)^{\prime}$. The simplest form of such
decomposition occurs when the $T_i$ are mutually orthogonal
projections. This type of decomposition corresponds to that
determined by othogonal measure.
So, taking the spectral resolution of density matrix $\varrho_{\omega}$
we obtain the very special (subcentral) orthogonal decomposition.
Therefore, if we restrict ourselves to decomposition induced by
spectral resolution of $\varrho_{\omega}$, in general, 
we can not expect to attain $\inf_{\mu \in 
M_{\omega}(\cS)} \int d\mu(\varphi) S(r \varphi)$, see also Subsection VC.

\subsection{Examples I}

To illustrate the question of computation of EoF we start with
very simple models.
\begin{enumerate}
\item  The von Neumann entropy (for finite systems)
is maximal for the state of the form
$\omega_{\varrho_m}(A) = Tr \varrho_m A$ with $\varrho_m = {1 \over dim{\cal H}} {\bf 1}$
($dim$ stands for dimension).  For such state it is equal to ${\rm ln}(dim {\cal H})$
and this is the maximal value of $ E$.
\item  Let us consider $2 \times 2$ system with $\cH_1 \equiv \cH_2$
( so $dim {\cal H}_1 = 2 = dim{\cal H}_2$)
and the singled state $\Psi_-$ defined as $|\Psi_-> = 
{1 \over {\sqrt2}}(|01> - |10>)$. Here we adopt a notation
of quantum mechanics by writing $|01> \equiv 
e_0 \otimes e_1$ where $\{ e_0, e_1 \}$ is a basis in $\cH_1$, etc.
Write $\omega_{\Psi_-}(A) =  Tr \{ |\Psi_-><\Psi_-|\cdot A \otimes B \}$ where $A \in 
{\cal B}({\cal H}_1)$ while $B \in {\cal B}({\cal H}_2)$.
Then $r \omega_{\Psi_-}(A) = Tr \{ |\Psi_-><\Psi_-|\cdot A \otimes {\bf 1} \} 
= Tr \{ ({1 \over 2}{\bf 1}) A \}$. So ${E}(\omega_{\Psi_-}) = {\rm ln} 2$. 
\item Let us consider $d \times d$ system and so called maximally
entangled state $|\Psi_+^d> = {1 \over {\sqrt{d}}} \sum_{i=1}^{d} |i>\otimes |i>$
where $\{|i>\}$ is a basis in ${\cal H}_1 = {\cal H} = {\cal H}_2$.
Again, let us define $\omega_{\Psi_+^d}(A \otimes B)= Tr \{ |\Psi_+^d><\Psi_+^d|
A \otimes B \}$ and consider $r\omega_{\Psi_+^d}$. It is easy to note
that $r\omega_{\Psi_+^d}(A) = Tr \{ ({1 \over d} {\bf 1}) A \}$.
Hence ${ E}(\omega_{\Psi_+^d}) = {\rm ln }d$, so
$ E$ attains its maximal value.
\end{enumerate}

The just listed results are easy to show since 
there is no question concerning the non-uniqueness
of decomposition of the (pure) state $\omega$ into pure states. 

\subsection{Convexity of EoF}

To prove convexity of EoF let us show that the set
$M_{\lambda_1 \omega_1 + \lambda_2 \omega_2}(\cS)$
contains the sum of the sets $\lambda_1 M_{\omega_1}(\cS)$
and $\lambda_2 M_{\omega_2}(\cS)$ where $\lambda_1$ and $\lambda_2$
are non-negative numbers such that $\lambda_1 + \lambda_2 =1$.
To see this we recall (see e.g. \cite{BR} or \cite{Alf})
that $\mu \in M_{\omega}(\cS)$ if and only if $\mu(f) \ge f(\omega)$
for any continuous, real-valued, convex function $f$.
Thus
\begin{equation}
(\lambda_1 \mu_1 +\lambda_2 \mu_2)(f) \ge \lambda_1 f(\omega_1)
+\lambda_2 f(\omega_2) \ge f(\lambda_1 \omega_1 + \lambda_2 \omega_2)
\end{equation}
implies the above stated relation between sets. 
Hence
\begin{eqnarray}
E(\lambda_1 \omega_1 + \lambda_2 \omega_2)
= \inf_{\mu \in M_{\lambda_1 \omega_1 + \lambda_2 \omega_2}(\cS)}
\int d\mu(\varphi) S(r\varphi) \nonumber \\ 
\le \lambda_1
\inf_{\mu \in M_{\omega_1}(\cS)} \int d\mu(\varphi) S(r \varphi) \nonumber \\
+ \lambda_2 \inf_{\mu \in M_{\omega_2}(\cS)} \int d\mu(\varphi)
S(r \varphi) 
 = \lambda_1 E(\omega_1) + \lambda_2 E(\omega_2)
\end{eqnarray}

Consequently, the function $\cS \ni \omega \mapsto E(\omega)$
is the convex one. 

\subsection{Subadditivity}
Consider the tensor product of von Neumann algebras
$\cB(\cH_1) \otimes \cM \otimes \cB(\cH_1) \otimes \cM$
and a state $\omega \otimes \omega$ over it where $\omega$ is a state on
$\cB(\cH_1) \otimes \cM$.
We observe
\begin{eqnarray}
E(\omega \otimes  \omega) =
\inf_{\mu \in M_{\omega \otimes \omega}(\cS_T)}
\int d\mu(\nu) S_{1+2}(r\nu)  \le \nonumber \\ 
\inf_{\mu_1 \times \mu_2 \in M_{\omega}(\cS) \times M_{\omega}(\cS)}
\int d\mu_1(\nu) \int d\mu_2(\nu^{\prime}) S_{1+2}(r
\circ \nu \otimes  \nu^{\prime})
\nonumber \\
\le \inf_{\mu_1 \times \mu_2 \in M_{\omega}(\cS) \times M_{\omega}(\cS)}
\int d\mu_1(\nu) \int d\mu_2(\nu^{\prime}) (S_1(r\nu) \nonumber \\
+ S_1(r\nu^{\prime})) = 2 E(\omega)
\end{eqnarray}

where $\cS_T$ denotes the set of all states on 
$\cB(\cH_1) \otimes \cM \otimes \cB(\cH_1) \otimes \cM$,
$S_{1+2}$ ($S_1$) the von Neumann entropy on $\cB(\cH_1)
\otimes \cB(\cH_1)$ ($\cB(\cH_1)$ respectively).
The last inequality follows from subadditivity of the von Neumann 
entropy.
Consequently, EoF has also a form of subadditivity. 
Applying the above argument to $E(\omega \otimes ...\otimes \omega)$
one can consider the "density" of EoF and treat $E(\omega)$ as an extensive
(thermodynamical) quantity. This feature of EoF seems to be important in quantum information (cf. \cite{Hor1}).

\subsection{Topological properties of EoF}

Now we wish to examine the question of continuity
of EoF. 
To describe that topological property we shall need some preliminaries.
Let us consider $M_1(\cS)$ and $\cS$ as two compact spaces
and a continuous mapping $b$ of $M_1(\cS)$ onto $\cS$
given by $M_1(\cS) \ni \mu \mapsto b(\mu) = \int_{\cS} \nu d\mu(\nu)$,
so $b(\mu)$ is the barycenter of the measure $\mu$.
Moreover, let us consider the equivalence relation $E(b)$
on the set $M_1(\cS)$ determined by the decomposition $\{ b^{-1}(\omega)
\}_{\omega \in \cS}$ of $M_1(\cS)$ into fibers of $b$.
We denote by $q$ the mapping of $M_1(\cS)$ to $M_1(\cS)/E(b)$
assigning to the point $\mu \in M_1(\cS)$ the equivalence class
$[\mu] \in M_1(\cS)/E(b)$.
We equip $M_1(\cS)/E(b)$ with the quotient topology, so $q$ is the natural 
(quotient) mapping.
As $b$ is a continuous mapping of the compact (Hausdorff) space $M_1(\cS)$
to the compact (Hausdorff) space $\cS$ then the equivalence relation $E(b)$ is closed.
 We wish to represent the mapping $b: M_1(\cS) \to \cS$ 
as the composition $\overline{b} \circ q$ of the natural mapping
$q$ with the mapping $\overline{b}$ of the quotient space
$M_1(\cS)/E(b)$ onto $\cS$ defined by letting $\overline{b}(b^{-1}(\omega))
= \omega$. It is an easy observation that the mapping
$\overline{b}$ is continuous. Hence we have

\newcommand{\End}{\operatorname{End}}
\setlength{\unitlength}{1mm}
\begin{center}
\begin{picture}(70,30)
\put(40,25){\makebox(0,0){$\cS$}}
\put(10,25){\makebox(0,0){$M_1(\cS)$}}
\put(25,5){\makebox(0,0){$M_1(\cS)/E(b)$}}
\put(18,25){\vector(1,0){15}}
\put(33,8){\vector(1,2){5}}
\put(12,18){\vector(1,-2){5}}
\put(24,27){\makebox(0,0){${\scriptstyle b}$}}
\put(12,14){\makebox(0,0){${\scriptstyle q}$}}
\put(38,14){\makebox(0,0){${\scriptstyle \overline{b}}$}} 
\end{picture}
\end{center}

In particular, $\overline{b}$ is one-to-one continuous mapping
of $M_1(\cS)/E(b)$ onto $\cS$. We want to show that $\overline{b}$ is 
a homeomorphism. To prove this 
we observe that $q$ is the continuous mapping carrying the compact topological space $M_1(\cS)$ onto topological space $M_1(\cS)/E(b)$. Then,
$M_1(\cS)/E(b)$ is pre-compact (not necessary Hausdorff space).
But, then $\overline{b}$ is the continuous mapping
of the pre-compact space $M_1(\cS)/E(b)$ onto compact (Hausdorff)
space $\cS$. Thus, $\overline{b}$ is a homeomorphism.
Therefore, we arrived to

\begin{prop}
$\overline{b}$ is a homeomorphism; i.e. the mapping $b$ is quotient.
\end{prop}

Now we wish  to describe equivalence classes in $M_1(\cS)/E(b)$. 
Let $\omega_{\alpha} \in \cS$. We observe

\begin{equation}
b^{-1}(\omega_{\alpha}) = \{ \mu \in M_1(\cS);
\int \nu d\mu(\nu) = \omega_{\alpha} \}
\end{equation}

In other words, $b^{-1}(\omega_{\alpha})$ is equal to the set
$M_{\omega_{\alpha}}(\cS)$ of all probabilistic measures
on $\cS$ which represent the point $\omega_{\alpha}  \subset \cS$.
We recall, see \cite{Alf}, that $\mu \in M_{\omega_{\alpha}}(\cS)$
is equivalent to $\mu \sim \delta_{\omega_{\alpha}}$, i.e.
that $\mu$ is equivalent to the Dirac measure $\delta_{\omega_{\alpha}}$
where the equivalence of (probabilistic) measures $\mu$ and $\mu^{\prime}$
is defined as
\begin{equation}
\int\nu d\mu(\nu) = \int \nu d\mu^{\prime}(\nu)
\end{equation}

But $\delta_{\omega_{\alpha}} \sim \mu$, in turn, is equivalent to 
(cf. \cite{Alf}) $\delta_{\omega_{\alpha}} - \mu \in \cN(\cS)$
where $\cN(\cS)$ is the annihilator of $A(\cS)$ in the dual pair
$<C_{\Rn}(\cS), M_{\Rn}(\cS)>$.
Here, $A(\cS)$ ($C_{\Rn}(\cS), M_{\Rn}(\cS)$)
is the set of all continuous real-valued affine functions on $\cS$
(the vector space of all real-valued continuous functions, the vector
space of all real measures on $\cS$ respectively).
Now it should be clear that an equivalence class $[\mu]$
in $M_1(\cS)/E(b)$ is equal to $\{ \delta_{\omega} + \mu;
\mu \in \cN(\cS), \mu(f) \ge -f(\omega)\quad for \quad any \quad
0 \le f \in C_{\Rn}(\cS) \}$ with some fixed $\omega \in \cS$.

In order to complete our discussion of the diagram we should examine topological 
properties of the set-valued map $b^{-1}$. 

\begin{prop}
The set-valued map $b^{-1}$ is upper semicontinuous.
\end{prop}
$Proof$:
The map $b^{-1}$ is lower
(upper) semicontinuous (cf. \cite{Engelking}) if and only if
for every closed set $K \subset M_1(\cS)$ the set
$\cO_L = \{ \omega: b^{-1}(\omega) \subset K \}$
(the set $\cO_U = \{ \omega: b^{-1}(\omega) \cap K \ne \emptyset \}$)
is closed in $M_1(\cS)$. 
To examine upper semicontinuity of $b^{-1}$ let us consider a net
$\{ \omega_{\alpha} \} \subset \cO_U$ with a limit $\omega_0$.
Futhermore, let us choose measures $\{ \mu_{\alpha} \in 
b^{-1}(\omega_{\alpha}) \cap K \}$. As $K$ is a compact subset there exists
a convergent subnet $\{ \mu_{\beta} \}$ such that
$\mu_{\beta} \to \mu \in K$. Since $\int_{\cS} \nu d\mu_{\beta}(\nu)
=\omega_{\beta}$ we infer that
\begin{equation}
\omega_{\beta} \to \omega = \int_{\cS} \nu d\mu(\nu)
\end{equation}
Then, the uniqueness of the limit implies
$\omega_0 = \omega$ and the proof of upper semicontinuity 
of $b^{-1}$ is complete.

\vskip 1cm

Having fully clarified topological relations among $\cS$,
$M_1(\cS)$, and $M_1(\cS)/E(b)$ we wish to prove

\begin{prop}
EoF, $\cS \ni \omega \mapsto E(\omega)$, is the continuous function.
\end{prop}
$Proof$: We start with an easy proof of lower 
semicontinuity of EoF. To this end let $\{ \omega_{\alpha} \}$
be a net with a limit $\omega$ and let $E(\omega_{\alpha}) \le s$, where
$s$ is a real number. To show that $E(\omega) \le s$ let us take
$\epsilon >0$ and choose $\mu_{\alpha} \in M_{\omega_{\alpha}}(\cS)$
such that $\mu_{\alpha}(S \circ r) < s + \epsilon$.
As $M_1(\cS)$ is a compact set there exists a convergent subnet
$\{ \mu_{\beta} \}$ with the limit $\mu_0$.
Let $\hat{A}$ be an affine, real-valued continuous function on $\cS$.
Then
\begin{equation}
\hat{A}(\omega_{\beta}) = \mu_{\beta}(\hat{A}) \to \mu_0(\hat{A})
\end{equation}
and $\hat{A}(\omega_{\beta}) \to \hat{A}(\omega)$. Thus
$\mu_0 \in M_{\omega}(\cS)$. Therefore, $s + \epsilon \ge
lim \mu_{\beta}(S \circ r) = \mu_0(S \circ r) \ge E(\omega)$.
Consequently, $E(\omega) \le s$ and the proof of lower
semicontinuity of EoF is complete.

Now, let us consider the question of upper
semicontinuity of $E(\omega)$. Again, let $\omega_{\alpha}$ be a net with 
a limit point $\omega_0$. Take $s$ such that $E(\omega_{\alpha}) \ge s$.
We observe that for any $\mu_{\alpha} \in M_{\omega_{\alpha}}(\cS)$,
$\mu_{\alpha}(S \circ r) \ge s$.
Again, the use of a convergent subnet $\{ \mu_{\beta} \}$
with a limit $\mu_0$ implies that 
$s \le \lim_{\beta} \mu_{\beta}(S \circ r)  = \mu_0(S \circ r)$,
where $\mu_0 \in M_{\omega_0}(\cS)$.
Thus, to prove upper upper semicontinuity of $E(\omega)$ it is enough to show 
that any $\mu \in M_{\omega_0}(\cS)$ can be obtain as a limit
of $\{ \mu_{\beta}^{\prime} \}$, i.e. 
$M_{\omega_0}(\cS) \ni \mu = \lim_{\beta} \mu_{\beta}^{\prime}$ where
$\mu_{\beta}^{\prime} \in M_{\omega_{\beta}}(\cS)$.
In particular, we want to have
\begin{equation}
\label{dwiegwiazdki}
\wedge_{\epsilon >0} \wedge_{\mu \in M_{\omega_0}(\cS)} \vee_{\mu_{\beta}
\in M_{\omega_{\beta}}} \quad |\mu(f) - \mu_{\beta}(f)| < \epsilon
\end{equation}
for a continuous function $f$ on $\cS$.
Assume contrary, i.e.,

\begin{equation}
\label{gwiazdka}
\vee_{\epsilon >0} \vee_{\mu \in M_{\omega_0}(\cS)} \wedge_{\mu_{\beta}
\in M_{\omega_{\beta}}}
\quad |\mu(f) - \mu_{\beta}(f)| \ge \epsilon
\end{equation}
for a continuous function $f$.
We note (cf. \cite{Engelking}) that the upper semicontinuity of $b^{-1}$ implies
that for every open set $U \subset M_1(\cS)$
the set $\{ \omega: b^{-1}(\omega) \subset U \}$
is open. Now, assuming (\ref{gwiazdka})
one can find  the neighbourhood $U_{\mu}$ of $\mu$
which does not contain any $\mu_{\beta}$.
But, $\{ \omega: b^{-1}(\omega) \subset U_{\mu} \}$
is a neighbourhood of $\omega_0$.
The convergence $\omega_{\beta} \to \omega_0$
implies that each neighbourhood of $\omega_0$ should contain
states of the form $\omega_{\beta}$. 
Thus, also, $\{ \omega: b^{-1}(\omega) \subset U_{\mu} \}$
should contains many $\omega_{\beta}$.
Hence, one can
find $M_{\omega_{\beta}}(\cS) \subset U_{\mu}$.
But this contadicts (\ref{gwiazdka}). Therefore, (\ref{dwiegwiazdki})
holds and the proof of upper
continuity of $E(\omega)$ is complete. This completes 
the proof of Proposition.

\vskip 1cm

As $\cS \ni \omega \mapsto E(\omega)$ is a continuous convex
function,  
the application of the Bauer maximum principle 
leads to:
\begin{cor}
$E(\omega)$ attains its maximum at an extremal point
of $\cS$, so the family of maximally entangled states is a subset
of pure states.
\end{cor}

\section{Examples and concluding remarks}

\subsection{Examples II}

Let us begin this final section with
 some other illustrative examples
showing the usefulness of EoF.

\smallskip

\begin{enumerate}
\item Let $\omega = \sum_k \omega_k$ be a decomposition of the state $\omega$.
Then, an application of convexity would lead to the following
estimation of entanglement of $\omega$:  $E(\omega) \le 
\max_k E(\omega_k)$.
\item In the discussion of entangled states 
the positive partial transposition criterion
plays an important role (\cite{hor}, \cite{per}). 
To consider that question, let us put $\cM = \cB(\cH_2)$
with finite dimensional Hilbert space $\cH_2$.
We recall that
the map $\alpha: {\cal B}({\cal H}_1 \otimes {\cal H}_2) \to
{\cal B}({\cal H}_1 \otimes {\cal H}_2)$ with $\alpha = id \otimes \tau$
where $\tau$ is a transposition map of the matrix representation of
an arbitrary $B \in 
{\cal B}({\cal H}_2)$
in a certain fixed basis, provides 
essential ingredient of that criterion.
Let us define $(\alpha^d \omega)(A \otimes B) 
\equiv \omega(\alpha (A \otimes B))$
and let us note
\begin{eqnarray}
(r\alpha^d\omega)(A) = (\alpha^d\omega)(A \otimes {\bf 1})
=\omega(\alpha(A \otimes {\bf 1}) \nonumber \\
 = \omega(A \otimes \tau({\bf 1}))
=\omega(A \otimes {\bf 1}) = (r\omega)(A)
\end{eqnarray}
Consequently, the partial transposition does not change the measure of entanglement.
Therefore, the basic point of that criterion is that $id \otimes \tau$ is not a completely
positive map. For further details on relations between entanglement and positive maps
see \cite{MM}.
\item Let us consider $d \times d$ system with the corresponding Hilbert space
${\cal H} \equiv \cH_1 \otimes \cH_2$ and let 
$P$ be a projector such that $P{\cal H}$ does not contain product states.
We recall that such projectors are related to the concept
of unextendible product bases \cite{ben}. Let us define the state
$\omega_P$ as $\omega_P(A \otimes B) = (Tr P)^{-1} Tr \{ P \cdot A \otimes B \}$.
We want to judge whether $\omega_P$ is a separable state. To this end
let us consider an arbitrary decomposition of $\omega_P$, i.e., 
\begin{eqnarray}
 \omega_P(A \otimes B) = (TrP)^{-1} Tr P A \otimes B = \\
 (Tr P)^{-1} \sum_k TrPa_k P A \otimes B \equiv
\sum_k \lambda_k \omega_k(A \otimes B)\nonumber
\end{eqnarray}
where operators $a_k \ge 0$ are defined on $\cH$ and satisfy
$\sum_k a_k = \bf 1$ while $\omega_k$ stands for the state
determined by $Pa_kP$. 
Assume $\omega_P$ is a separable state. 
Then, there is a decomposition $\omega_P = \sum_k \lambda_k \omega_k$
such that $r\omega_k$ is a pure state for any $k$.
But, then $\omega_k$ would be a product state, i.e.
\begin{equation}
\omega_k(A \otimes B) = Tr \varrho_k^1 \otimes \varrho_k^2 A \otimes B
\end{equation}
This would imply
\begin{equation}
\varrho_k^1 \otimes \varrho_k^2 = constant \cdot P a_k P
\end{equation} 
But, this is impossible as $P\cH$ does not contain product (vector) states.
Consequently,
${E}(\omega_P) >0$.
\end{enumerate}

\subsection{Comparison with the Bennett, DiVincenzo, Smolin and Wooters 
definition of EoF}
 The presented examples with explicitly calculated $E(\omega)$ 
suggest a close relation of our EoF with 
the original definition of EoF, given by Bennett {\it et al},
which will be denoted by $EoF_B$. Obviously, $EoF \le EoF_B$. To examine
the converse inequality
we start with a simple observation that
\begin{eqnarray}
\inf_{\mu \in M_{\omega}(\cS)} \int d\mu(\nu) S(r\nu)
= \inf\{ \sum_{i=1}^n \lambda_i S(r\nu_i):    \\
\omega = \sum_{i=1}^n \lambda_i \nu_i \quad(convex\quad sum) \}
\nonumber\\
\end{eqnarray}
and the first infimum is attained for some $\mu \in M_{\omega}(\cS)$. 
The above 
observation follows from the fact that each measure $\mu$ can be ($^*$weakly) 
approximated by measures with finite support. 
On the other hand, measures concentrated on $\cS_p$,
where $\cS_p$ is the set of all pure states, are known to be
maximal with respect to the order $\mu \prec \nu$ ($\mu \prec \nu$
if and only if $\mu(f) \le \nu(f)$ for any convex, 
real-valued convex function $f$,
cf. \cite{phel} or \cite{Alf}), so minimal on the set of all concave
functions. It particular, such the measure
is minimal on $S\circ r$. Apparently, the maximality of a measure 
(on all convex functions) is too strong demand and therefore 
to get the converse inequality between EoF and $EoF_B$ some extra argument
should be given (see the next subsection).

\subsection{Optimal decompositions}

In this section we wish to examine the question
of existence of the very special type of decompositions, so called 
optimal decompositions.
A decomposition $\omega = \sum_j \lambda_j \varrho_j$, where
$\{ \varrho_i \}$ are pure states, such that the 
infimum in the definition of EoF is attained will be called 
{\it an optimal decomposition}. In other words, the infimum
is attained by a measure $\mu_0$ with finite support
contained in the set of all pure states. Thus
$$E(\omega) = \inf_{\mu \in M_{\omega}(\cal S)}\int_{\cal S}
S(r \varrho) d\mu(\varrho) = \int_{\cal S} d\mu_0(\varrho) S(r \varrho)$$
with $supp \mu_0 = \{ \varrho_1,..., \varrho_n \}$, $n< \infty$
and $\varrho_i \in \cS_p$.
Clearly, $\mu_0 = \sum_1^n \lambda_i \delta_{\varrho_i}$
where $\delta_{\varrho}$ stands for the Dirac measure, $\{ \varrho_i \}$ are 
pure states and $\omega = \sum \lambda_i\varrho_i$.
We recall that each measure has maximal balayage and 
each state $\omega \in \cal S$ is the barycenter of a measure which
is maximal for the order $\succ$ (for all necessary details see
 \cite{BR}, \cite{phel}, \cite{Alf} or \cite{Mey}). 
It is an easy observation that 
$\mu_0$ is a maximal measure.

{\it In order to avoid misunderstaning we repeat: 
entropy is concave, maximality is defined for convex functions,
so inf for "$x \mapsto - xlnx$" means sup for "$x \mapsto xlnx$"}.

We recall that maximality of the measure $\mu_0$ implies:
\begin{equation}
\label{etykieta}
\mu_0 \quad \textrm{ is supported
 by the set } \{ \omega; \omega \in \cS, f(\omega)
= \overline{f}(\omega) \}
\end{equation}
 for all continuous convex functions on $\cS$
where $\overline{f}$ is the upper envelope of $f$
i.e., $\overline{f}(\omega) = inf \{ g(\omega);
-g \in P({\cal S}), g \ge f \}$,
($P(\cS)$ stands for the set of continuous convex functions on $\cS$).
Moreover, if $\overline{f}$ is the upper envelope of
a continuous function on $\cS$ then
\begin{equation}
\label{etykietaI}
\overline{f}(\omega) = \sup_{\mu \in M_{\omega}(\cS)} \int f(\nu) d\mu(\nu)
\end{equation}

Denote the set of all maximal measures in $M_{\omega}(\cS)$
by $\cZ_{\omega}(\cS)$. One can show (cf. \cite{Alf}) that
$\cZ_{\omega}(\cS)$ is a face of $M_{\omega}(\cS)$. The set of
all extremal measures in $\cZ_{\omega}(\cS)$ will be denoted by
$\cE^Z_{\omega}(\cS)$, i.e.
\begin{equation}
\cE^Z_{\omega}(\cS) = \cE_{\omega}(\cS) \cap \cZ_{\omega}(\cS)
\end{equation}
where $\cE_{\omega}(\cS)$ stands for simplicial measures (cf. Section IV).
Let us consider (cf. \cite{Alf})
\begin{equation}
F_0 = \{ \mu \in M_{\omega}(\cS); \mu(f) =
\overline{f}(\omega) \}
\end{equation}
where $f$ is a convex continuous function on $\cS$.
Obviously, by virtue of (\ref{etykietaI}) the set $F_0$ is not empty.
Let $\mu \in F_0$ and assume $\mu = \lambda_1 \mu_1 + (1 - \lambda_1) \mu_2$
where $\mu_1, \mu_2 \in M_{\omega}(\cS)$
and $\lambda_1 \in (0,1)$. Clearly, $\mu_1$ and $\mu_2$ should be in $F_0$.
Thus, $F_0$ is a face. It is an easy observation that $F_0$
is the closed set with the property:
$\mu \in F_0$, $\nu \in M_{\omega}(\cS)$ and $\mu \prec \nu$
imply $\nu \in F_0$, i.e., $F_0$ is the hereditary upwards face.
Then, the application of the Lumer existence theorem (see 
Proposition 1.6.4 in \cite{Alf}) proves
existence of a measure $\mu_0$ in $F_0 \cap \cE^Z_{\omega}(\cS)$.
Thus, we proved:
\begin{prop}
Maximum of the set $\{ \mu(-S \circ r); \mu \in M_{\omega}(\cS) \}$
 for a continuous convex function $-S$ is attained
by a simplicial boundary measure.
\end{prop}
\vskip 0.5cm
Now, let us examine very special case. Namely,
if the Hilbert space $\cH$ of the composite system is 
of the finite dimension,
say, $dim\cH = n$ then $\cS$ can be considered as a compact subset in
the space $\Rn^{(2n)^2}$. On the orther hand, a probability measure
$\mu$ on a convex compact subset in $\Rn^{(2n)^2}$ is simplicial 
if and only if $\mu$ is supported by affinely independed set of
(at most $(2n)^2 +1)$ points. 
Moreover, in the considered (finite dimensional) case
the weak-$^*$ topology is metrizable. Hence, maximal measures
are supported by extremal points in $\cS$. Combining the just given
results for the concave function
$S \circ r$ we can infer the existence
of optimal decomposition. Obviously, it does not exclude a possibility that
$infimum$ is attainable on some other measures.
Nevertheless, for finite
dimensional case we have: $EoF = EoF_B$.

Turning to others decompositions (cf. Section IV.A) we note
that even in two dimensional case (so for the algebra
of $2 \times 2$ matrices) one can write simplicial maximal measure that is 
not orthogonal (cf. \cite{BR}). Therefore, even for low dimensional 
case one can not say that EoF, $E(\omega)= \inf_{\mu \in \cO_{\omega}(\cS)}
\mu(S \circ r)$.

Finally we want to point out that the 
proof existence of optimal
decomposition depends only on the structure of the set
$M_{\omega}(\cS)$.

\subsection{Remarks}
There is a frequently considered question of uniqueness of a measure of 
entanglement. We already observed, see comments following definition
EoF in Section III, that one can replace the von Neumann entropy by any
continuous non-negative
concave function which vanishes on pure states, e.g.
$\varrho \mapsto Tr\{ \varrho(\bf{1} - \varrho) \}$.
Then, all argument can be repeated and we would arrive to a new measure
of entanglement.
As a matter of fact the only reason for our assumption of 
$dim \cH_1 < \infty$ was that in this case, the von Neumann entropy
is continuous. Thus, to perform further generalization of EoF one can replace
$\varrho \mapsto -Tr\{ \varrho ln \varrho \}$ by another function from the 
just mentioned class. Clearly, changing the von Neumann entropy function,
the argument leading to the subadditivity of $E(\omega)$ should be modified.
The additional reason to use the von Neumann entropy throughout the paper
follows from the fact that
there is a nice relation (cf. \cite{Con}, \cite{CNT}, \cite{BNU})
between the entropy $H$  of subalgebra $\cM_1$ in $\cM_1 \otimes \cM$
relative to the state $\omega$ ($\cM_1$, and $\cM$ are von Neumann 
algebras), the von Neumann entropy of the restricted state and EoF:
\begin{equation}
H_{\omega, \cM_1 \otimes \cM}(\cM_1) = S(r\omega) - E(\omega)
\end{equation}
It is clear that the main difficulty for calculating $H$ is encoded
in $E(\omega)$. Therefore, our results concerning EoF shed some new light on
 the nature of $H$.

\vskip 1cm
It is a pleasure to thank Karl Gerd H. Vollbrecht for fruitful
comments and criticisms and to Marcin Marciniak for
many helpful discussions. We would also like to thank Micha{\l}
Horodecki for helpful comments.
This work has been supported by KBN grant PB/0273/PO3/99/16  


\end{document}